\documentclass[aps,prd,twocolumn,preprintnumbers,nofootinbib,superscriptaddress,amsmath]{revtex4-2}
\usepackage[english]{babel}
\usepackage[colorinlistoftodos, color=green!40, prependcaption]{todonotes}
\usepackage[pdftex, pdftitle={Article}, pdfauthor={Author}]{hyperref} 
\usepackage{dcolumn}
\usepackage{bm}
\usepackage{multirow}

\usepackage{amsmath}
\usepackage{amsfonts,amssymb}
\usepackage{hyperref}
\usepackage{graphicx}
\usepackage{natbib}
\usepackage[utf8]{inputenc}
\usepackage{xcolor}
\usepackage{bm}
\usepackage{diagbox}
\usepackage{hhline}
\usepackage[normalem]{ulem}

\usepackage[normalem]{ulem}

\usepackage{xcolor}

\begin{document}

\title{A novel approach to cosmological non-linearities as an effective fluid}

\author{Leonardo Giani}
\email{l.giani@uq.edu.au}
\affiliation{School of Mathematics and Physics, The University of Queensland, Brisbane, QLD 4072, Australia}
\author{Rodrigo von Marttens}
\affiliation{Instituto de Física, Universidade Federal da Bahia, 40210-340, Salvador, BA, Brazil}
\affiliation{PPGCosmo, Universidade Federal do Espírito Santo, Vitória-ES, 29075-910, Brazil}
\author{Ryan Camilleri}
\affiliation{School of Mathematics and Physics, The University of Queensland, Brisbane, QLD 4072, Australia}

\begin{abstract}
We propose a two parameters extension of the flat $\Lambda$CDM model to capture the impact of matter inhomogeneities on our cosmological inference. Non virialized but non-linearly evolving overdense and underdense regions, whose abundance is quantified using the Press-Schechter formalism, are collectively described by two effective perfect fluids $\rho_{\rm{c}},\rho_{\rm{v}}$ with non vanishing equation of state parameters $w_{\rm{c,v}}\neq 0$. 
These fluids are coupled to the pressureless dust, akin to an interacting DM-DE scenario. The resulting phenomenology is very rich, and could potentially address a number of inconsistencies of the standard model, including a simultaneous resolution of the Hubble and $\sigma_8$ tensions. To assess the viability of the model, we set initial conditions compatible to the Planck 2018 best fit $\Lambda$CDM cosmology and fit its additional parameters using SN~Ia observations from DESY5, BAO distances from DESI DR2 and a sample of uncorrelated $f\sigma_8$ measurements. Our findings show that backreaction effects from the cosmic web could restore the concordance between early and late Universe cosmological probes.

\end{abstract}

\maketitle

\section{Introduction} 
Within the concordance model $\Lambda$CDM, Cosmic Microwave Background (CMB) observations \cite{Planck:2018vyg} suggest that the cosmic web of Large Scale Structures (LSS) we observe in the Universe emerged from small perturbations of an otherwise homogeneous, isotropic and spatially flat background. 
Huge progress has been made towards a better understanding of the cosmic web and its evolution, in particular through numerical simulations in both a Newtonian~\cite{Jenkins:2000bv,Teyssier:2001cp,Springel:2005nw,Vogelsberger:2014dza} and relativistic context~\cite{Adamek:2016zes,Adamek:2020jmr,Macpherson:2021gbh,Magnall:2023tzm,Williams:2024vlv,Barrera-Hinojosa:2019mzo,Quintana-Miranda:2023eyn}.

However, the impact of LSS on the global evolution of the Universe, usually referred to as cosmological backreaction~\cite{Sicka:1999cb,Ellis:2005uz,Marra:2007pm,Green:2010qy,Buchert:2011sx,Bolejko:2017lai,Koksbang:2019glb,Buchert:2019mvq,Gasperini:2009mu,Gasperini:2009wp,Gasperini:2011us,Schwarz:2010px}, is still a matter of debate. 
It is expected that averaging on sufficiently large scales the familiar FLRW metric is recovered, with backreaction described by some effective modification of the Friedmann equations. Different approaches lead to different conclusions, with a spectrum ranging from no appreciable impact ~\cite{Ishibashi:2005sj,Green:2014aga,Green:2016cwo,Macpherson:2018btl} to backreaction being responsible for the (apparent) accelerated expansion of the Universe ~\cite{Rasanen:2003fy,Alnes:2005rw,Biswas:2006ub,Buchert:2007ik,Wiltshire:2009db,Heinesen:2020sre,Racz:2016rss} (see also Refs.~\cite{Fleury:2013sna,Buchert:2015iva,Bolejko:2015gmk,Giani:2023aor,Anton:2023icm,Clifton:2024mdy,Ben-Dayan:2012uam,Bagheri:2014gwa,Rubart:2014lia,Ben-Dayan:2014swa,Buchert:2002ij} for examples of the intermediate spectrum, and Ref.~\cite{Kolb:2009rp} for a proposed classification of backreaction models).
Another possibility ~\cite{Baumann:2010tm} is to model the impact of the non-linearities using the formalism of Effective Field Theory (EFT) of LSS \cite{Carrasco:2012cv,Lewandowski:2016yce}.

At present, there seems to be a consensus \cite{Buchert:1995fz,Peebles_2010,Marra:2011ct, Sanghai:2015wia} that backreaction from virialized structures is negligible. However, a non-negligible impact from non-virialized structures with characteristic scale $\ell\geq 1 \;\rm{Mpc}$ in our cosmological inference cannot be excluded, as supported by a number of investigations performing averages of cosmological observables in exact inhomogeneous spacetimes \cite{Clifton:2019cep,Clarkson:2011zq,Clifton:2019cep} and relativistic simulations \cite{Adamek:2018rru,Adamek:2014gva,Adamek:2017mzb}. These results support the idea that backreaction should vanish if averages are performed in constant time hyphersurfaces in the Newtonian Gauge, whose existence and connection with real observers is still unclear.

In this letter, we propose a novel approach to model these averages of the non-linearities and inhomogeneities as an effective fluid, and explore its impact on cosmological observables. The resulting phenomenology has interesting implications for a number of inconsistencies of the $\Lambda$CDM model, including the Hubble and $\sigma_8$ tensions \cite{Abdalla:2022yfr}, and the preference for Dynamical Dark Energy (DDE) hinted by Sn~Ia and BAO measurements from DES \cite{DES:2024tys} and DESI \cite{DESI:2024mwx,DESI:2025zgx} when combined with CMB observations.

\section{Effective fluid description
}

Our starting point is the assumption that the 
total matter density $\rho_{\rm{tot}}$ consist of dust ($\rho_{\rm{d}}$) and two effective fluids ($\rho_{\rm{c}}$ and $\rho_{\rm{v}}$) describing backreaction effects from clustering regions and expanding voids. We couple these effective fluids to the dust, since they emerge at late times as a result of the gravitational interactions between dust fluid elements. Under these hypotheses, the Friedmann equations takes the form
\begin{equation}\label{fe}
    3H^2 = \rho_{\rm{tot}} +\Lambda\;,
\end{equation}
where $\rho_{\rm{tot}}= \rho_{\rm{d}}+\rho_{\rm{c}} + \rho_{\rm{v}}$. The continuity equations become
\begin{eqnarray}
    &\dot{\rho}_{\rm{d}}& + 3H\rho_{\rm{d}} = -C(t) - V(t) \;, \label{dce} \\ 
    &\dot{\rho}_{\rm{v}}& + 3H\rho_{\rm{v}}\left(1+w_{\rm{v}}\right) = V(t) \;, \label{vce} \\
    &\dot{\rho}_{\rm{c}}& + 3H\rho_{\rm{c}}\left(1+w_{\rm{c}}\right) = C(t) \;, \label{cce}
\end{eqnarray}
with $w_{\rm{c}}$,$w_{\rm{v}}$ denoting the Equation of State (EoS) parameters of $\rho_{\rm{c}},\rho_{\rm{v}}$, a dot the derivative w.r.t. the cosmic time, and $C(t)$ and $V(t)$ the coupling between the matter species. Notice that $\rho_{\rm{tot}}$ is conserved
\begin{equation}\label{tce}
    \dot{\rho}_{\rm{tot}} + 3H\rho_{\rm{tot}}\left(1 + w_{\rm{tot}}\right)=0\;,
\end{equation}
and has EoS parameter
\begin{equation}\label{wtwcwv}
w_{\rm{tot}}=\frac{\rho_{\rm c}}{\rho_{\rm tot}}w_{\rm c} + \frac{\rho_{\rm v}}{\rho_{\rm tot}}w_{\rm v}\;.
\end{equation}
Eqs.~\eqref{fe},~\eqref{tce} and \eqref{wtwcwv} can be mapped into the scalar-averaged backreaction equations from Buchert \cite{Buchert:2019mvq} by defining,
\begin{eqnarray}\label{buchertmap}
    \mathcal{Q}\equiv -\frac{1}{2}\left[\Delta\rho +3 \left(w_{\rm{c}}\rho_{\rm{c}}+w_{\rm{v}}\rho_{\rm{v}}\right)\right]\;,\\
    ^3\mathcal{R}\equiv \frac{3}{2}\left[-\Delta\rho + \left(w_{\rm{c}}\rho_{\rm{c}}+w_{\rm{v}}\rho_{\rm{v}}\right)\right]\;,
\end{eqnarray}
where $\mathcal{Q}$ and $^3\mathcal{R}$ are the kinematic backreaction and the averaged spatial curvature, $H$ becomes the averaged expansion rate $\theta$ of the volume $\mathcal{V}$, and where we have defined $\Delta\rho \equiv \rho_{\rm{tot}}-\bar{\rho}$, with $\bar{\rho}\propto \mathcal{V}^{-1}$. It can be easily shown that the closure relation
\begin{equation}
    \dot{\mathcal{Q}} + \;^3\dot{\mathcal{R}} + 2\theta\left( \;^3\mathcal{R} +3\mathcal{Q} \right)=0\;,
\end{equation}
reduces to the condition \eqref{wtwcwv}. Our effective model can be therefore interpreted as describing the evolution of volume averages in a inhomogeneous universe containing non-linearly expanding and collapsing objects.
In the remainder of this section we provide physically motivated ansatz for the functional forms of $C(t),V(t)$ and $w_{\rm{tot}}$, and explore their cosmological implications. 

\subsection{Abundance of non-linear, non-virialized objects}

We need to quantify the amount of regions in the Universe relevant for backreaction effects.
Since virialized structures should not contribute any, we focus on structures evolving non-linearly that haven't virialized yet. 
To estimate their abundance, we make use of the Press–Schechter (PS) formalism, whose theory and generalizations have been developed e.g. in Refs.~ \cite{Press:1973iz,1991ApJ...379..440B,Sheth:2003py,Jennings:2013nsa,Verza:2024rbm}.
The spherical collapse model \cite{1967ApJ...147..859P,1972ApJ...176....1G} predicts that the linear density contrast threshold $\delta$ for the turnaround to occur is $\delta_{ta}\approx 1.06$ and for virialization $\delta_{c}\approx 1.68$.\footnote{The impact of anisotropies on the gravitational collapse and the PS formalism has been considered for example in Refs.~\cite{Sheth:1999su,Angrick:2010qg,DelPopolo:2018zsf,DelPopolo:2021tjc,Giani:2021gbs}, whereas the impact of different dark energy models has been extensively studied for example in Refs.~\cite{Mota:2004pa,Pace:2010sn,Bartelmann:2005fc,Meyer:2012nw,Tarrant:2011qe,Pace:2017qxv,Batista:2021uhb}} According to the PS formalism, the integrated probability for a smoothed density perturbation $\delta_R$ to be within the range $\delta_{ta}\leq\delta_R\leq\delta_{c}$ is
\begin{equation}\label{fraction_collapsing}
f_{\rm{c}}\left(\delta_{ta}\leq\delta_R \leq\delta_{c}\right)\equiv f_{\rm{PS}}(\delta_{\rm{ta}}) - f_{\rm{PS}}(\delta_{\rm{c}})\;,
\end{equation}
where $f_{\rm{PS}}$ is given by 
\begin{equation}\label{PSfraction}
    f_{\rm{PS}}\left(>\delta_c\right) = 1-\rm{erf}\left(\frac{\delta_c}{\sqrt{2}\sigma_R}\right)\;.
\end{equation}
We consider voids to be non-linear when their extrapolated linear density contrast becomes smaller than $\delta_{\rm{nl}}<-1$, whereas virialization occurs at $\delta_{v}=-2.7$ \cite{Jennings:2013nsa}. The integrated probability in this case is given by
\begin{equation}\label{fraction_voids}
    f_{\rm{v}}(\delta_{v} \leq \delta_{R} \leq \delta_{\rm{nl}})\equiv f_{\rm{void}}(\delta_{\rm{nl}})-f_{\rm{void}}(\delta_{v})\;,
\end{equation}
with $f_{\rm{void}}$ defined as \cite{Jennings:2013nsa}
\begin{equation}
    f_{\rm{void}}\equiv \int_{\tilde{R}}^\infty \frac{dR}{R}\;\frac{df_{\rm{ln}\sigma}}{dR}\frac{d \ln \sigma^{-1}}{d\ln{R}}\;, \label{fractionvoids}
\end{equation}
where the fractional density of voids is
\begin{equation}\label{Jenknumdens}
 \frac{df_{\rm{ln}\sigma}}{dR}=
 \begin{cases}
     \sqrt{\frac{2}{\pi}}\frac{|\delta_v^{\rm{lin}}|}{\sigma} e^{-\left(\frac{\delta_v^{\rm{lin}}}{\sqrt{2}\sigma}\right)^2}  \quad\quad\quad \rm{for}\;\; x\leq 0.276\;,\\
     2\sum_{j=1}^{4}e^{-\frac{(j\pi x)^2}{2}}j\pi x^2\sin\left(j\pi \mathcal{D}\right)\;\; \rm{for}\;\; x > 0.276 \;.
 \end{cases}
 \end{equation}

Using the above definitions, we formulate the following ansatz for the energy densities entering Eq.~\eqref{fe}
\begin{equation}\label{ansatz1}
\rho_{\rm{d}}=\rho_{\rm{tot}}\left(1- f_{\rm{v}} - f_{\rm{c}}\right)\;, \quad \rho_{\rm{c}}=\rho_{\rm{tot}}f_{\rm{c}}\;, 
\quad \rho_{\rm{v}}=\rho_{\rm{tot}}f_{\rm{v}}\;. 
\end{equation}

\subsection{The source terms $C(t)$ and $V(t)$}
Given the abundances in Eq.~\eqref{ansatz1},  the background dynamics of model is fully specified by the coupling functions $C\left(t\right)$, $V\left(t\right)$ and the EoS parameters $w_{\rm{c}}$ and $w_{\rm{v}}$.
The former appear as source terms in the continuity equations, and can be derived from Eq.~\eqref{ansatz1} combined with Eqs.~\eqref{dce} and \eqref{cce}. It is convenient to combine both in a single interacting term $Q\left(t\right)\equiv C\left(t\right)+V\left(t\right)$ given by 

\begin{equation} \label{ansatz1bis}
   {Q} =\left[\dot{f}_{\rm c} + \dot{f}_{\rm v} + 3Hw_{\rm{tot}}\left(1-f_{\rm v} - f_{\rm c}\right)\right]\rho_{\rm{tot}}\;.
\end{equation}

\subsection{Spherical collapse ansatz for $w_{\rm{tot}}$}
Our next assumption is that $\rho_{\rm{tot}}$ can be written as
\begin{equation}\label{energybudg}
    c^2\rho_{\rm{tot}}= c^2\left(\rho_{\rm{dust}} +\rho_{\rm{nl}}\right) \equiv \frac{M_{\rm{dust}}c^2 + M_{\rm{nl}}c^2+ \sum_i E_{i}^{\;\rm{nl}}}{V}\;,
\end{equation}
where we have defined the total masses contained in dust and non-linear structures $M_{\rm{dust}}$ and $ M_{\rm{nl}}$, and the internal energies of the latter $E_{i}^{\rm{nl}}$. 

Following the spirit of the PS formalism, we compute $E_i$ using the spherical collapse model. The curvature $K_i$ associated to a spherical region of initial radius $R(t_0)$, density perturbation $\delta(t_0)$ and total mass $M_i$ is given by \cite{Weinberg:1987dv,Weinberg:2008zzc}
\begin{equation}
    c^2 K_i = \frac{10}{3} M_i G \frac{\delta (t_0)}{R(t_0)}\;,
\end{equation}
to which we associate an internal energy
\begin{equation}\label{Esphcollapse}
    E_i =-\frac{3}{8\pi G} \frac{c^4 K_i}{R^2}\frac{V_i}{R^2_i}  =- \frac{5}{3}M_ic^2 \delta(t)\frac{R_i(t)}{\bar{R}(t)}\;,
\end{equation}
where $V_i$ is the volume of the sphere and $\bar{R}(t)$ the radius that it would have if it was expanding with the background. 

Our definition Eq.~\eqref{Esphcollapse} implies that Eq.~\eqref{energybudg} becomes the volume average of the expansion rates of the background and of the non-linearly evolving spherical regions. 
To proceed further, we have to define the terms $M_{\rm{nl}}$, $M_{\rm{dust}}$ appearing in Eq.~\eqref{energybudg}. A sensible choice is 
\begin{equation}\label{massi}
    M_{\rm{nl}}\equiv \sum_i M_i \;, \qquad M_i=\frac{4\pi}{3}\rho_{\rm{tot}}R^{3}_i\;,
\end{equation} 
where we have defined $M_i$ as the integral of  $\rho_{\rm{tot}}$ within a spherical volume.\footnote{Another possibility is to use $\rho_{\rm{dust}}$ in  the definition~\eqref{massi}, i.e. define the mass of these objects as the total amount of dust within the region. The differences between these two are of second order in $f_i,\dot{f}_i$ and the results qualitatively unchanged.}
Since $\rho_{\rm{tot}}$ is not pressureless, the mass in these objects is non-conserved, $\dot{M}_i \neq 0$.  
Using Eqs.~\eqref{PSfraction},\eqref{fractionvoids} we can write
\begin{equation}\label{Mnltot}
    \frac{M_{\rm{nl}}}{V}\equiv \int_{\tilde{R}_c}^\infty dR\; \frac{df_{c}}{dR} \frac{M_i}{V_i} + \int_{\tilde{R}_v}^\infty dR\; \frac{df_{v}}{dR} \frac{M_i}{V_i} = \rho_{\rm{tot}}\left(f_v+f_c\right)\;, 
\end{equation}
where the phenomenological parameters $\tilde{R}_c$ and $\tilde{R}_v$ fix the minimum scale of collapsing and expanding regions contributing to backreaction effects. 
Notice that the above implies that the total mass in dust should be identified with
\begin{equation}
    \frac{M_{\rm{dust}}}{V}\equiv \rho_{\rm{tot}}-\frac{M_{\rm{nl}}}{V} -\frac{\sum_iE_i}{Vc^2}\;.
\end{equation}

Taking the time derivative of Eq.~\eqref{energybudg} and assuming conservation of the total mass, $\dot{M}_{\rm{tot}}=0$, Eq.~\eqref{tce} gives
\begin{equation}
w_{\rm{tot}}= -\frac{1}{3H\rho_{\rm{tot}}}\frac{\sum_i \dot{E}_i}{V}\;.    
\end{equation}
Notice that for $E_i=0$ we have $\frac{M_{\rm{nl}}}{M_{\rm{dust}}} = \frac{\rho_{\rm{nl}}}{\rho_{\rm{dust}}}$ and $w_{\rm{tot}}=0$, in agreement with the fact that if $\rho_{\rm{nl}}$ has no internal energy, then there is no difference between the total energy density in rest mass and $\rho_{\rm{tot}}c^2$.\footnote{This, ultimately, is why we use $\rho_{\rm{tot}}$ rather than $\rho_{\rm{dust}}$ in the mass definition Eq.~\eqref{massi}.} 

A direct consequence of Eqs.~\eqref{ansatz1} and \eqref{ansatz1bis} is that the EoS parameter $w_{\rm{tot}}$ depends non-trivially on $\rho_{\rm{c}},\rho_{\rm{v}}$ and their time evolution, and therefore $\tilde{R}_v$, $\tilde{R}_c$. Notice that $f_{\rm v}$ and $f_{\rm c}$ decay exponentially for $R_i \gg 1$, recovering $\Lambda$CDM in this limit. 

\subsection{Impact of the effective fluid on the growth of perturbations}

Eqs.~\eqref{fe}-\eqref{cce} effectively describe an interacting DM-DE model with $\rho_{\rm{DE}}=\Lambda + \rho_{\rm{c}}+\rho_{\rm{v}}$. Since $\rho_{\rm{c}}$ and $\rho_{\rm{v}}$ are purely phenomenological descriptions of dust interactions, they have no physical meaning at a fluid-element level. We therefore assume that, just like $\Lambda$, they do not cluster.\footnote{Hence we neglect on sub-horizon scales the impact of merging non-virialized clusters and voids.} Including the interaction term, the perturbation equations for the dust fluid in the Newtonian gauge and the sub-horizon limit become \cite{2010JCAP...09..029L}:
\begin{eqnarray}
    \dot{\delta}+\theta=\frac{Q}{\rho_{\rm tot}}\left(\delta-\frac{\delta Q}{Q}\right)\,, \label{pert_energy}\\
    \dot{\theta}+H\theta+\Delta\phi=0\,, \label{pert_momentum}
\end{eqnarray}
where $\delta$ is the density contrast, $\theta$ is the divergence of the spatial 3-velocity, $\phi$ is the gravitational potential and $\delta Q$ the perturbation of the coupling term.
As it is customary, we also neglect perturbations of the Hubble rate in the sub-horizon limit~\cite{vonMarttens:2018iav}.
Linearizing Eq.~\eqref{ansatz1bis} and neglecting perturbations of the term within the square brackets,\footnote{This approximation is justified by the expectation that both $f_{\rm v}$ and $f_{\rm c}$, along with their respective derivatives, are small.} one obtains $\delta Q \approx Q\delta$.
Under these hypotheses, we obtain
\begin{equation}\label{deltaeq}
    \ddot{\delta} + 2H\dot{\delta} - 4\pi G \rho_{\rm{d}} \delta  = 0\;.
\end{equation}
Although Eq.~\eqref{deltaeq} retains the same form as in $\Lambda$CDM, the evolution of perturbations is now influenced by the quantities $H$ and $\rho_{\rm{d}} = \rho_{\rm{tot}}\left(1 - f_{\rm{v}} - f_{\rm{c}}\right)$. Therefore, to understand the behaviour of the model, we need to solve numerically the system of Eqs. \eqref{fe},\eqref{tce} and \eqref{deltaeq} under the hypotheses $\eqref{ansatz1}$, \eqref{energybudg}, \eqref{massi} and \eqref{Mnltot}.

\section{Numerical analysis}
Eqs.~\eqref{fe},\eqref{tce} predict the same expansion history as a $\Lambda$CDM cosmology until relatively low-redshift. We numerically integrate them together with Eq.~\eqref{deltaeq} using the Boltzmann solver CLASS \cite{2011arXiv1104.2932L} to obtain the initial conditions at redshift $z_{i}=20$, and assume that the transfer function is unchanged.\footnote{A Jupyter notebook with the code we use for the numerical integration is available at \href{https://github.com/Leolardo/Effective-fluids-Backreaction}{https://github.com/Leolardo/Effective-fluids-Backreaction}.} Unless otherwise stated, all our plots are obtained by fixing the initial conditions at $z_i$ to the flat $\Lambda$CDM best fit (TT,TE,EE+lowE) from Planck 2018(PL18) \cite{Planck:2018vyg}.\footnote{$\omega_b=0.0223828, \Omega_m = 0.3166$, $\theta^s_{100}=1.04090$, $A_s=2.101549 \times 10^{-9}$, $n_s = 0.9660499$ and $\tau_{re}=0.05430842$.}  


Due to their relevance for the recently suggested evidence for DDE, we fit the free parameters $\tilde{R}_{\rm{c}}$, $\tilde{R}_{\rm{v}}$ using the latest catalogues of SN~Ia from the Dark Energy Survey supernova program (DESY5) \cite{DES:2024tys} and BAO distance measurements from \cite{DESI:2025zgx}.  We also use a sample of uncorrelated measurements of $f\sigma_8$ from the latest Sloan Digital Sky Survey Baryon Oscillation Spectroscopic Survey data release (SDSS IV) \cite{eBOSS:2020yzd} and peculiar velocity measurements from 6df and SDSS \cite{Said:2020epb,Lai:2022sgp}.

For the SN~Ia we use the \textit{Cobaya} \citep{cobaya2, cobaya1}\footnote{\url{https://github.com/CobayaSampler/cobaya}} MCMC sampler,\footnote{The convergence of MCMC chains was assessed in terms of a generalized version of the $R-1$ Gelman-Rubin statistic. We adopt a more stringent tolerance than \textit{Cobaya}'s default value of $R-1=0.001$.} and for $f\sigma_8$ and BAO measurements we use \textit{emcee} \cite{emcee}.\footnote{We assess the numerical convergence following the prescription in
https://emcee.readthedocs.io/en/stable/user/autocorr/}. We adopt flat log priors $0<\log{\tilde{R}_i}< 1.5$, in such a way that we only consider backreaction effects from inhomogeneities larger than $\tilde{R}_i\geq 1$ Mpc/h. The upper bound is such that deviations from $\Lambda$CDM are below $<0.01 \%$. 

\begin{figure}[]
    \includegraphics[scale=0.75]{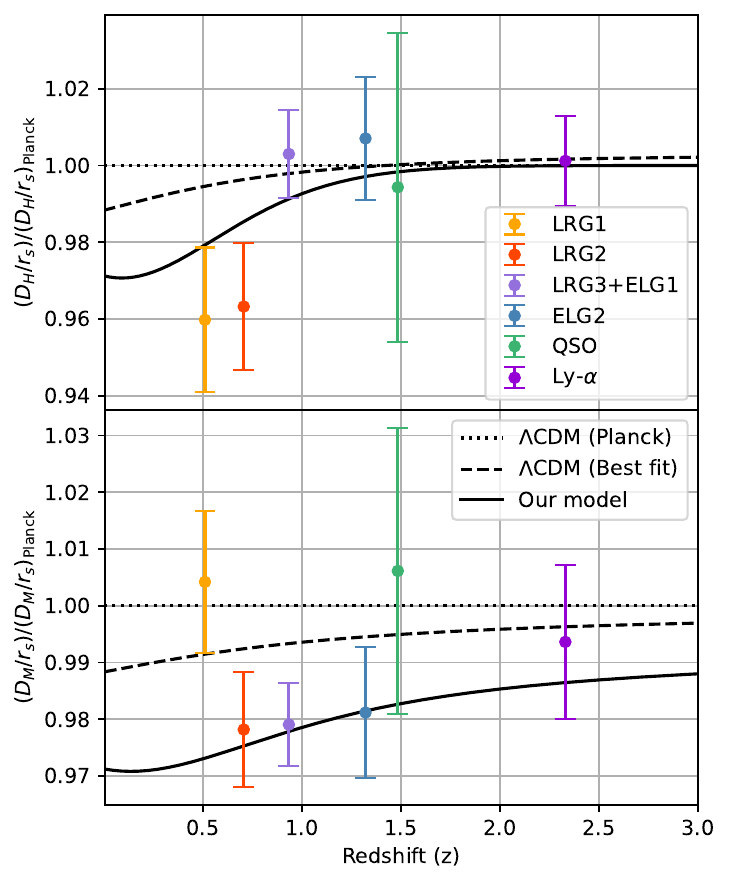}\;\\ 
	\caption{The evolution of the distances $D_H$ and their integral $D_M$ in our model and in $\Lambda$CDM for the Planck and the DESI best fits. As clear from table ~\ref{modelbestfits}, our model provides a remarkably good best fit to BAO data over $\Lambda$CDM, comparable with the CPL best fit from Ref.\cite{DESI:2025zgx}.}
 \label{BAO_distances}
\end{figure}

\begin{figure}[]
    \includegraphics[scale=0.75]{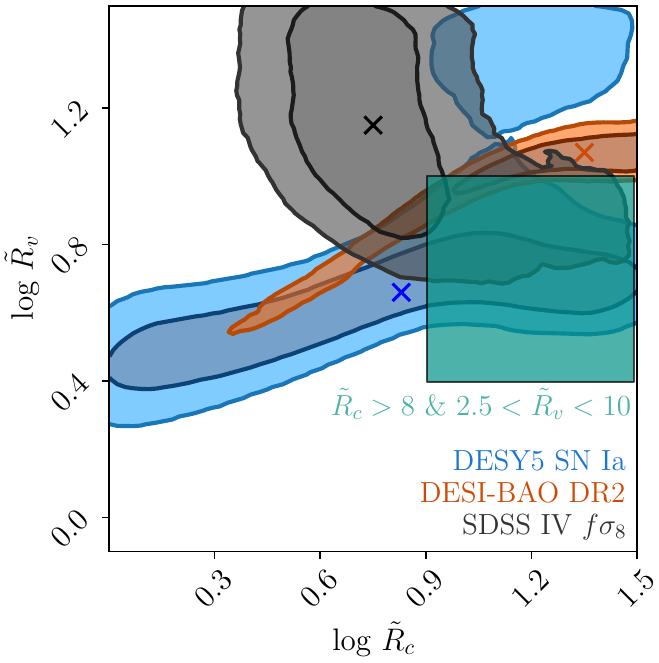}\;\\ 
	\caption{Constraints on the free parameters $\tilde{R}_c,\tilde{R}_v$ for DESI DR2 BAO(orange),  DESY5 SnIa (blue) and $f\sigma_8$ (black) measurements  from \cite{eBOSS:2020yzd,Said:2020epb,Lai:2022sgp}. The green box indicates a region of parameter space that could alleviate both the $H_0$ and $\sigma_8$ tensions (see Fig.~\ref{H0-S8splot}).} 
 \label{data_constraints}
\end{figure}

\begin{figure}[]\hspace*{-1cm}
    \includegraphics[scale=0.55]{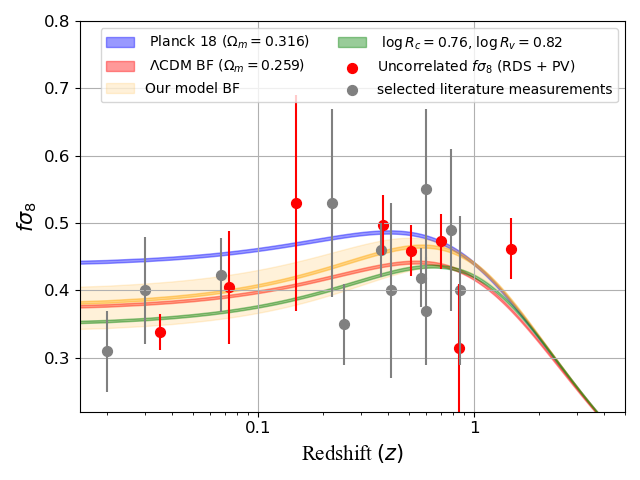}
	\caption{ Predictions for the evolution of $f\sigma_8$ from Pl18 (blue), $\Lambda$CDM best fit (red), our model's best fit (orange) and with $\tilde{R}_{\rm{v}}$ fixed to the DESY5 peak (green). Red dots indicate the uncorrelated $f\sigma_8$ measurements used for the fit \cite{eBOSS:2020yzd,Lai:2022sgp,Said:2020epb} and the grey points a collection of correlated measurements from the recent literature \cite{Davis:2010sw,Qin:2019axr,BOSS:2013uda,Lilow:2021omg,Pezzotta:2016gbo,2011MNRAS.415.2876B,BOSS:2013yzh,2012MNRAS.423.3430B}.}
 \label{fsigma8}
\end{figure}

Fig.~\ref{data_constraints} shows our exploration of the 2-dimensional parameter space $\tilde{R}_{\rm{c}},\tilde{R}_{\rm{v}}$. Table~\ref{modelbestfits} shows the $\chi^2$ values obtained for the $\Lambda$CDM, the PL18 best fit and our model. Figs.~\ref{BAO_distances} and~\ref{fsigma8} illustrate the evolution of the BAO distances and of $f\sigma_8$ in these models.

We stress that these results are illustrative, and a more rigorous analysis should consider both datasets together with CMB data varying all the cosmological parameters. Our simple exercise, however, shows the potential of the model to address a few challenges of the $\Lambda$CDM.
\begin{table}[h]
\caption{The results of our MCMC parameter exploration of DESY5 and uncorrelated $f\sigma_8$ measurements}
\renewcommand{\arraystretch}{1.2}
\tiny
\centering
\begin{tabular}{c|ccc|cc}
\hline\hline
\multicolumn{1}{l|}{\multirow{2}{*}{}}                                                     & \multicolumn{3}{c|}{$\chi^2$}                                                                & \multicolumn{2}{c}{Parameters}                                                            \\
\multicolumn{1}{c|}{}                                                                      & \multicolumn{1}{l}{PL18} & \multicolumn{1}{l}{$\Lambda$CDM} & \multicolumn{1}{r|}{Our model} & \multicolumn{1}{l}{log $\tilde{R}_{\rm c}$} & \multicolumn{1}{l}{log $\tilde{R}_{\rm v}$} \\ \hline
\multirow{2}{*}{\begin{tabular}[c]{@{}c@{}}DESY5 SN\\ (1735$\ d.o.f$)\end{tabular}} & \multirow{2}{*}{1653.88} & \multirow{2}{*}{1649.88}         & \multirow{2}{*}{1648.22}       & \multirow{2}{*}{$0.83^{+0.43}_{-0.58}$}     & \multirow{2}{*}{$0.66^{+0.15}_{-0.16}$}     \\ 
                                                                                           &                          &                                  &                                &                                             &                                             \\ \hline
\multirow{2}{*}{\begin{tabular}[c]{@{}c@{}}DESI BAO (DR2)\\ (11$\ d.o.f$)\end{tabular}} & \multirow{2}{*}{17.85} & \multirow{2}{*}{17.17}         & \multirow{2}{*}{6.41}       & \multirow{2}{*}{$<1.00$}     & \multirow{2}{*}{$1.02^{+0.04}_{-0.07}$}     \\ 
                                                                                           &                          &                                  &                                &                                             &                                             \\ \hline
\multirow{2}{*}{\begin{tabular}[c]{@{}c@{}}RSD + PV\\ (8 $d.o.f$)\end{tabular}}            & \multirow{2}{*}{21.9}    & \multirow{2}{*}{11.7}            & \multirow{2}{*}{11.4}          & \multirow{2}{*}{$0.75^{+0.19}_{-0.15}$}     & \multirow{2}{*}{$1.15^{+0.24}_{-0.22}$}     \\
                                                                                           &                          &                                  &                                &                                             &                                             \\ \hline\hline
\end{tabular}
\label{modelbestfits}
\end{table}

\begin{figure*}[]
\centering
    \includegraphics[scale=0.4]{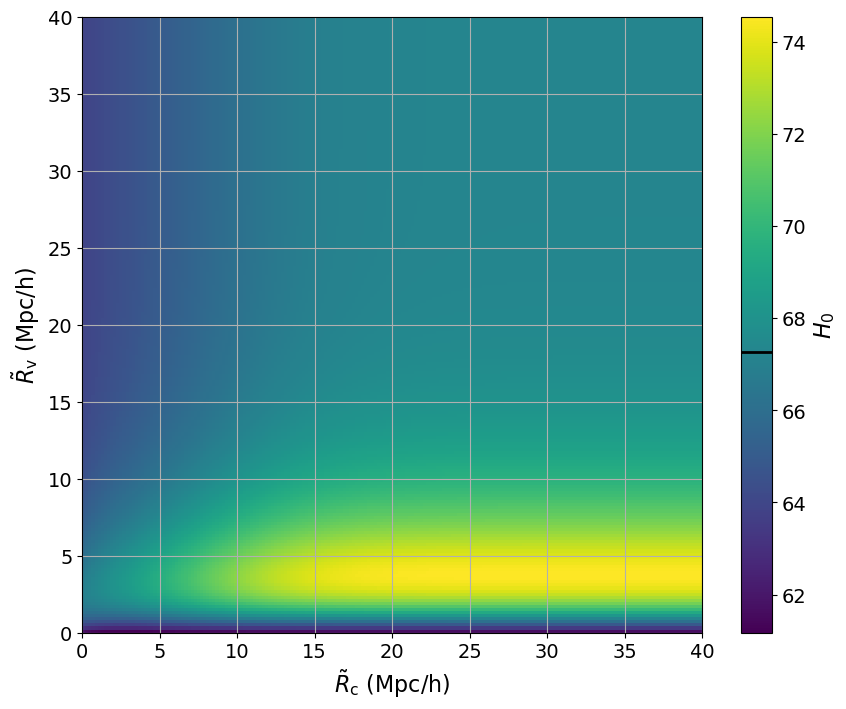}\;\includegraphics[scale=0.4]{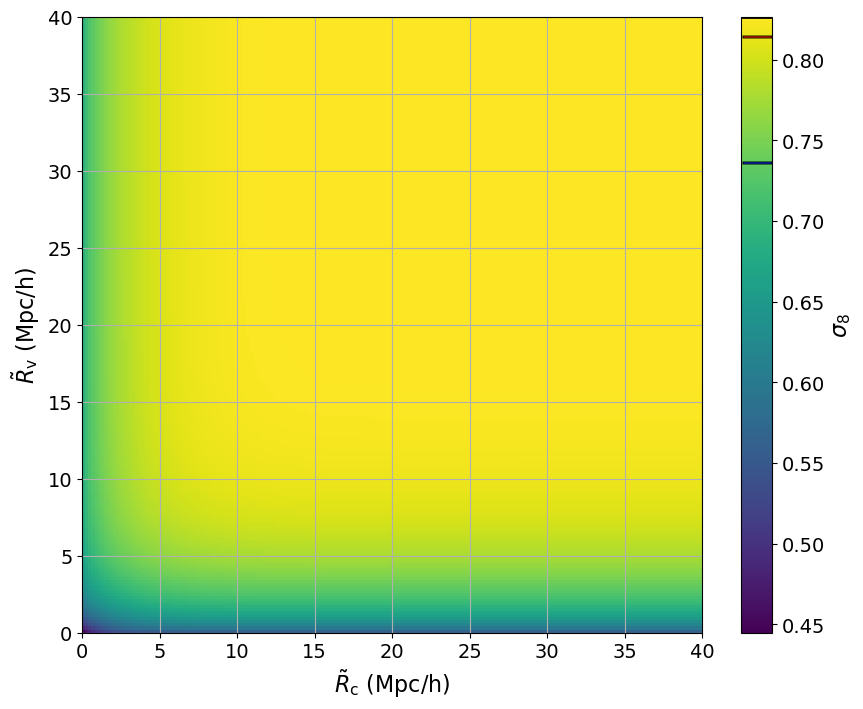}\\
	\caption{Variations of $H_0$ (left) and $\sigma_8 (z=0)$ (right) as a function of $\tilde{R}_c, \tilde{R}_v$. Interestingly, there is a region of the parameter space (around $\tilde{R}_c > 8$ and $2.5<\tilde{R}_v<7.5$ Mpc/h) that could alleviate both tensions at the same time. In the $H_0$ panel, the black line on the color bar denotes the PL18 $\Lambda$CDM best fit. The corresponding one in the $\sigma_8$ panel is at the top of the color bar. The red and blue lines in the $\sigma_8$ color bar consist to the results of DES Y3 $3\times 2$pt \cite{DES:2021vln} and KiDS-Legacy cosmic shear \cite{Wright:2025xka}.} 
 \label{H0-S8splot}
\end{figure*}

\section{Discussion}\label{conclusion}

This letter proposes a new prescription to compute the cosmological backreaction of matter inhomogeneities inspired by models of interacting DM-DE. We introduce a coupling between the matter non-linearities and the ``dust'' at the background level, describing the conversion of matter fluctuations into LSS. It is worth stressing that our proposal is not a Newtonian backreaction model. Indeed, as shown in Eqs.~\eqref{buchertmap}, the \textit{Buchertian} averaged spatial curvature $^3\mathcal{R}$ is in general non-vanishing.

Similarly to other backreaction proposals \cite{Wiltshire:2009db}, our model depends on a somewhat arbitrary choice of averaging scales. Common sense suggests that sufficiently small regions should have no impact on the background evolution of the Universe. This simple argument is backed up by several investigations in the literature \cite{Buchert:1995fz,Marra:2011ct,Baumann:2010tm,Kaiser:2017hqn}, which have shown that backreaction effects from virialized objects in a Newtonian framework are completely negligible. For this reason, we focus our attention to non-virialized regions, quantified using the PS formalism, with linear scales $\geq 1\; \rm{Mpc/h}$. We leave otherwise free the additional parameters of the model, $\tilde{R}_c$ and $\tilde{R}_v$, parametrising the minimum size of the inhomogeneities relevant for backreaction effects.  

Compared to $\Lambda$CDM, the model features a different expansion history at relatively low redshift, depending on the relative abundance of collapsing and expanding non-linear regions. The non-linear effective fluid is, by its own nature, a transient phenomena, since non-linearities will eventually virialize. If their minimum size is sufficiently small, their decay ($\dot{f}_i\leq<0$) leads to a switch in the sign of $C(t) + V(t)$ in Eq.~\eqref{ansatz1bis} and of the EoS parameter $w_{\rm{tot}}$ at $z\ll1$.

The recent indications for DDE \cite{DESI:2024mwx, DES:2024tys} occur when simultaneously combining SnIa, BAO and CMB information. In a nutshell, the disagreement between matter density predictions from high- and low-redshift probes\footnote{It has recently been suggested \cite{Colgain:2024ksa,Colgain:2024xqj} that a non-standard evolution of the matter fluid might already be present in these datasets individually.} drives the preference for DDE, which can reconcile the data thanks to its additional free parameters. When addressed through CPL parametrization, this DDE seems to imply the crossing of the phantom divide $w=-1$. For this reason, a number of work explored modifications of the matter equation of state parameter $w_{m}\neq 0$ \cite{Giani:2025hhs,Kumar:2025etf,Chen:2025wwn,Yang:2025ume}. We notice that the model presented here provides a natural fundamental intepretation of these proposals. As shown in Table~\ref{modelbestfits}, the model provides excellent fit to all the datasets considered in this work, comparable in all cases with their $\Lambda$CDM best fit, but fixing the initial conditions by construction to the PL18 best fit.
Remarkably, the allowed parameter space contains a region addressing both the $H_0$ and $\sigma_8$ tensions. Furthermore, the model makes clear predictions on the evolution of perturbative observables, such as the growth rate $f$. We expect that future measurements of these quantities at intermediate redshifts will provide valuable constraints on the allowed parameter space. 

Summarizing, this work proposes an innovative two-parameters extension of the $\Lambda$CDM equivalent to an interacting DM-DE scenario that aims to describe backreaction effects from the cosmic web. The parametrization is phenomenological, but fundamentally motivated by established ingredients of our theoretical understanding of structure formation, the Press-Schechter formalism and the spherical collapse model. The arising phenomenology has the potential to address a few major challenges of the concordance model whilst providing insightful clues to the ``fitting problem'' in Cosmology~\cite{Ellis:1987zz}.

\section{Acknowledgments}
It is a pleasure to thank Timothy Clifton, Tamara Davis, Asta Heinesen, Cullan Howlett, Valerio Marra, Oliver Piattella, Khaled Said and Sunny Vagnozzi for their useful comments and suggestions. 
LG and RC acknowledge the support of an Australian Research Council Australian Laureate Fellowship (FL180100168) funded by the Australian Government. RvM is suported by Funda\c{c}\~ao de Amparo \`a Pesquisa do Estado da Bahia (FAPESB) grant TO APP0039/2023.

\bibliographystyle{unsrturl}
\bibliography{stiff.bib}

\end{document}